\documentclass[runningheads]{llncs}
\usepackage{graphicx}
\usepackage{comment}
\newcommand{\etal}{\textit{et al}. }
\usepackage{bm}
\usepackage{multirow}
\usepackage{booktabs}
\usepackage{subfigure}
\usepackage{bm}
\newcommand\blfootnote[1]{%
  \begingroup
  \renewcommand\thefootnote{}\footnote{#1}%
  \addtocounter{footnote}{-1}%
  \endgroup
}
\begin{document}
\title{Motion Pyramid Networks for Accurate and Efficient Cardiac Motion Estimation}
%
\titlerunning{Motion Pyramid Networks}
%
\author{Hanchao Yu$^{\star}$ \inst{2}, Xiao Chen \inst{1}, Humphrey Shi$^{\dagger}$ \inst{3}, \\
Terrence Chen \inst{1}, Thomas S. Huang \inst{2}, Shanhui Sun$^{\dagger}$ \inst{1}
} 

\authorrunning{H. Yu et al.}

\institute{United Imaging Intelligence, Cambridge, MA \and University of Illinois at Urbana-Champaign \and University of Oregon \\
\email{$^{\dagger}${shihonghui3@gmail.com}, $^{\dagger}${shanhui.sun@united-imaging.com}}}

\maketitle              
\blfootnote{$^{\star}$ This work was carried out during the internship of the author at United Imaging Intelligence, Cambridge, MA 02140.}
\begin{abstract}
Cardiac motion estimation plays a key role in MRI cardiac feature tracking and function assessment such as myocardium strain. In this paper, we propose Motion Pyramid Networks, a novel deep learning-based approach for accurate and efficient cardiac motion estimation. We predict and fuse a pyramid of motion fields from multiple scales of feature representations to generate a more refined motion field. We then use a novel cyclic teacher-student training strategy to make the inference end-to-end and further improve the tracking performance. Our teacher model provides more accurate motion estimation as supervision through progressive motion compensations. Our student model learns from the teacher model to estimate motion in a single step while maintaining accuracy. The teacher-student knowledge distillation is performed in a cyclic way for a further performance boost. Our proposed method outperforms a strong baseline model on two public available clinical datasets significantly, evaluated by a variety of metrics and the inference time.
New evaluation metrics are also proposed to represent errors in a clinically meaningful manner.

\keywords{motion pyramid network  \and motion compensation \and cyclic knowledge distillation}
\end{abstract}

\section{Introduction}
Cardiac motion estimation in cardiac MRI (CMR) is one of the fundamental techniques for cardiac feature tracking (CMR-FT). In the feature tracking system, key points from manual annotation or automatic generation are initialized on one image and then tracked through time. Once the spatiotemporal locations of each point are known, clinical indices such as myocardium strain can be computed to assess the dynamic deformation functionality of the heart, which are more sensitive and earlier indicators of contractile dysfunction, compared with the frequently used ejection function (EF) \cite{hor2011magnetic}. Besides, motions can also help other tasks in CMR image analyses like reconstruction \cite{huang2019dynamic,seegoolam2019exploiting} and segmentation \cite{qin2018joint,zheng2019explainable,yang2019deep}. Since the ground-truth of motion is difficult to acquire, most of the deep learning-based works formulate it as an unsupervised learning problem. Under this setting, the searching space is large and the optimal is not unique due to the lack of ground truth motion field. In \cite{qin2018joint,zheng2019explainable}, motion field smoothness is used as constraints, at the cost of compromised estimation of large motions. To evaluate the tracker's performance, DICE coefficients, surface distance and endpoint error are often used in recent researches \cite{qin2018joint,zheng2019explainable,wang2019gradient}. Considering the clinical applications like myocardium strain which are computed along specific directions, these metrics are not well aligned with clinical interest.  

To address the cardiac feature tracking challenges, we propose a motion pyramid network, which predicts and fuses a pyramid of motion fields from multiple scales of feature representations to produce a refined motion field (Section~\ref{backbone}). We utilize a novel cyclic teacher-student training strategy to further improve the tracking performance (Section~\ref{cycle}). The teacher model is trained via progressive motion compensations in an iterative manner in order to handle large motion as well as minimizing smoothness constraint limitation (Section~\ref{compensation}). The student model learns from the teacher model and matches its accuracy performing a more efficient inference (Section~\ref{cycle}). To align with clinical interest, we propose a novel evaluation method where the error vector is decomposed into radial and circumferential directions. The extensive experiments demonstrate that our method outperforms the strong single-scale baseline model and a conventional deformable registration method on two public datasets using a variety of metrics.

\section{Related Work} 
Many recent works show promising results in the area of deep learning-based cardiac motion estimation and myocardium tracking \cite{qin2018joint,zheng2019explainable,krebs2019learning,yu2020foal}. In \cite{qin2018joint}, cardiac motion estimation and cardiac segmentation are formulated as a multi-task problem with shared feature encoder and independent task heads. In \cite{zheng2019explainable}, a U-net like the apparent-flow network is proposed with a semi-supervised learning framework. These methods use the smoothness constraint to keep feasible anatomy. 

There are also non-learning based methods that can be divided into two categories: optical flow-based and registration based.  \cite{wang2019gradient} shows the most recent gradient flow-based method. A Lagrangian displacement field based post-processing is used to reduce the end-point error, which may not be time-efficient. Image registration based methods~\cite{puyol2018fully,rueckert1999nonrigid,de2012temporal,shen2005consistent,shi2012comprehensive,tobon2013benchmarking,krebs2019learning,yu2019novel} were applied to solve cardiac motion estimation. Rueckert \etal~\cite{rueckert1999nonrigid} proposed a free form deformation (FFD) method for non-rigid registration and is applied to cardiac motion estimation recently \cite{shen2005consistent,shi2012comprehensive,tobon2013benchmarking,puyol2018fully,vigneault2017feature}. However, deformable registration methods require iterative searching in a large parameter space.

The teacher-student and knowledge distillation mechanisms are popular in computer vision and medical image analysis society \cite{kong2018invasive,liu2019selflow}. Kong \etal \cite{kong2018invasive} utilizes a $L_2$ loss on feature maps between the teacher and the student models to guide the training of the compressed student network. 

Pyramid processing is widely accepted in the computer vision community \cite{sun2018pwc,ranjan2017optical,yu2017computed,mei2020pyramid,xu2020deep}. In PwcNet \cite{sun2018pwc} the input of each level of the pyramid is the output of the last level. In contrast, our network estimates the motion independently from each level and generate a motion pyramid to perform the multi-level fusion.

\section{Motion Pyramid Networks} 
We propose Motion Pyramid Networks (MPN) for the cardiac motion tracking problem. Our method consists of a multi-scale motion pyramid architecture for motion prediction, progressive motion compensation for post-processing and a cyclic knowledge distillation strategy to learn the motion compensation and speed up inference. 
\subsection{Multi-scale motion pyramid} 
\label{backbone}
\begin{figure}[bt]
\centering
\includegraphics[width=1.0\textwidth]{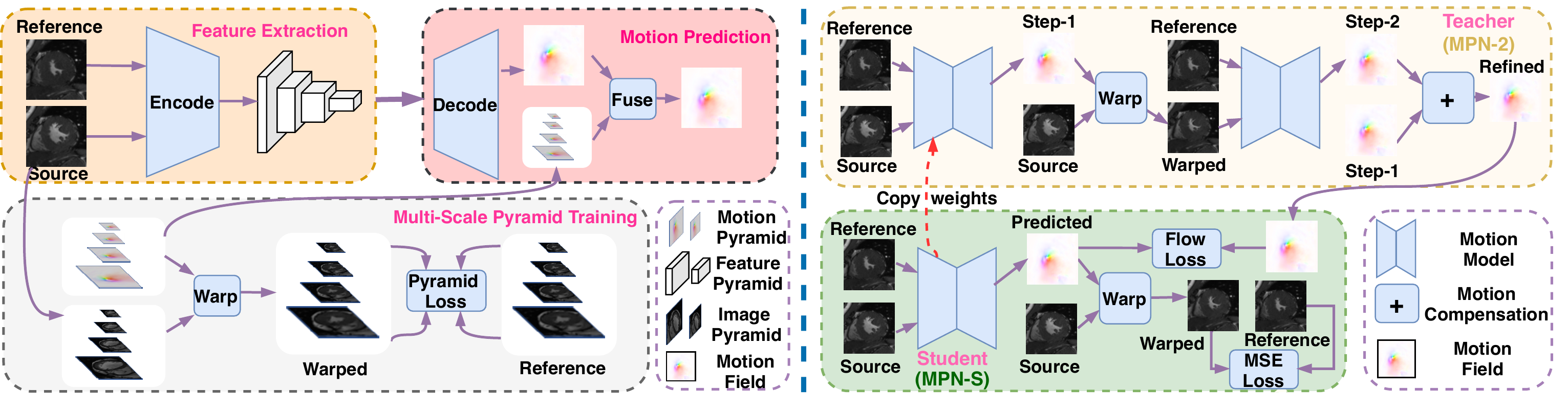}
\caption{Left: Overview of the proposed motion pyramid network (MPN). At each level of extracted features, the motion field is predicted. The decoder fuses the multi-scale features and generates a motion field which is further fused with the motion field pyramid to predict the refined motion field. Right: Illustration of the cyclic knowledge distillation method (MPN-C). The teacher model provides supervision through two-step compensation (MPN-2). The student model learns to estimate the motion under the guidance of the teacher as well as the same smoothness and MSE loss (MPN-S, w/o red dashed arrow). The weights of the student model are copied to the teacher (red dashed arrow) to initialize the next round of training a new student model (MPN-C).} \label{fig:overview}
\end{figure}

Fig.~\ref{fig:overview}(left) illustrates the overview of the proposed motion pyramid network (MPN). The features are extracted from source and target image independently using a shared encoder. We build a feature pyramid: the different scale features from the encoder are upsampled to the same size and concatenated channel-wise. The feature pyramid is then fed to the decoder which is used to predict an initial motion field. In addition, we build a motion pyramid as following: at each feature scale-level, a corresponding scale motion field is generated. Then the motion fields are upsampled to the same size and fused with the initial motion field through a fusion module to predict the refined motion. Note that we multiply a scaling factor to each motion field in the pyramid before fusion to make the motion fields comparable across scales, following PWCnet~\cite{sun2018pwc}. Furthermore, we leverage a deep supervision strategy to train the motion pyramid: the source image is downsampled to corresponding scales. The downsampled images are warped with the predicted motion pyramid to generate a warped image pyramid. The total loss is composed of the MSE loss between the warped and the reference image and the second-order smoothness loss \cite{qin2018joint}. Losses at each level $l$ are summed up as in Eq.~ (\ref{loss}).

\begin{equation}
    L_{total} = \sum_{l} L^{(l)}_{MSE} + \sum_{l} \lambda L^{(l)}_{smooth}
    \label{loss}
\end{equation}

\subsection{Progressive motion compensation}
\label{compensation}
Under certain smoothness constraint ($L_{smooth}$), it is a non-trivial task for the tracking learner to find a global optimal solution for all the cardiac motion variations even with pyramid utilization. This is because the smoothness constraint limits the searching space of the learner. Relaxing the smoothness can expand the searching space but the tracker becomes sensitive to disturbances such as noises and abrupt intensity changes. Nevertheless, we utilize some smoothness to maintain a feasible cardiac shape. However, the tracking accuracy drops in large motion cases. To solve this problem, we utilize a progressive motion compensation approach (MPN-2) at inference stage. Consider a model $M$ that takes frame $I_A$ and $I_B$ as input. Due to the aforementioned smoothness constraint problem, the warped image $I_C$ is an intermediate result between $I_A$ and $I_B$ and $M(I_A, I_B)$ actually yields motion field ${f}_{AC}$. Let $M(I_C, I_B)=f_{CB}$. Suppose $\bm{x_0}=(x, y)$ is a pixel in $I_A$ and $\bm{x_1, x_2}$ are the corresponding pixels in $I_C,I_B$, respectively. We have the following equations: $\bm{x_1} = f_{AC}(\bm{x_0}) + \bm{x_0}, \bm{x_2} = f_{CB}(\bm{x_1}) + \bm{x_1}$. Replace $\bm{x_1}$ with $\bm{x_0}$ in the second equation and notice that $\bm{x_2} - \bm{x_0} = f_{AB}(\bm{x_0})$, we can get $f_{AB}(\bm{x_0}) = f_{AC}(\bm{x_0}) + f_{CB}(f_{AC}(\bm{x_0})+ \bm{x_0})$, where $f_{AC}, f_{CB}, f_{AB}$ is step-1, step-2 and the refined motion field.
The derivation is based on forward warping (use $f_{AB}$ to warp $I_A$) while the same equation still holds with the backward warping  (use $f_{AB}$ to warp $I_B$) function. In forward warping, some pixels on the source are not on the target grids, leaving the target image with holes. In backward warping, each location in the target image is determined using backward flow to find the intensity in the source image via image interpolation. Moreover, the backward warping is differentiable and the gradients can be back-propagated, which was proved in [16]. In principle, this process could be multiple-step while in our work, we found that two-step progressive motion compensation is sufficient to solve the discussed problem.
\subsection{Cyclic knowledge distillation}
\label{cycle}
The used motion compensation method improves accuracy but increases inference time due to multi-step inferences. To solve the inference time problem, we introduced another model (a student), which is a replica of the trained tracker and further learns the knowledge from the progressive motion compensation steps (a teacher). We coined this teacher-student training method as MPN-S. The model (see Section \ref{backbone}) used in the teacher is coined as $M_t$ and the student $M_s$. $M_s$ is initialized using $M_t$. Parameters of $M_t$ are fixed in the teacher-student training step. Note that $M_t$ is used to generate both step-1 and step-2 motion fields. The refined flow from the teacher is used to supervise training the student $M_s$ utilizing a loss function $L_{flow} = ||f_{AB}^{t} - f_{AB}^{s}||_2$. Besides, $L_{MSE}$ is added as a self-supervision for the student model. Thus, we have the following total loss:
\begin{equation}
    L_{total} = L_{flow} + \mu L_{MSE} + \gamma L_{smooth}.
\end{equation}
Through the teacher's supervision, the student model attains the teacher's inference capability (two-step motion compensation). We further improve the student's inference capability via cyclic training strategy (MPN-C): when the current teacher-student training converges, the student model takes the teacher's role and a new round of teacher-student training is started. The student's performance can be continuously improved through this self-taught learning method. The overview of the process is shown in Fig.~\ref{fig:overview}(right).

\section{Experiments and Results}
\begin{figure}[bt]
\centering
\subfigure{
\includegraphics[width=0.18\linewidth]{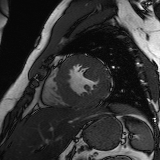}}
\subfigure{
\includegraphics[width=0.18\linewidth]{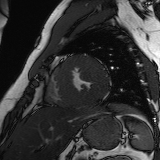}}
\subfigure{
\includegraphics[width=0.18\linewidth]{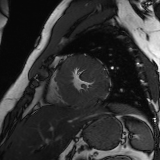}}
\subfigure{
\includegraphics[width=0.18\linewidth]{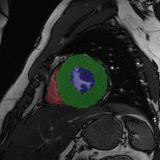}}
\subfigure{
\includegraphics[width=0.18\linewidth]{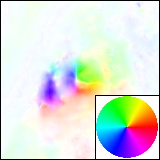}}
\subfigure{
\label{ED} 
\includegraphics[width=0.18\linewidth]{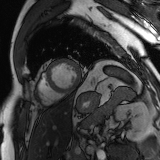}}
\subfigure{
\label{ES} 
\includegraphics[width=0.18\linewidth]{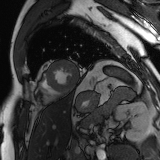}}
\subfigure{
\label{Warped} 
\includegraphics[width=0.18\linewidth]{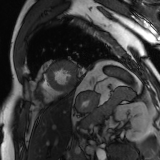}}
\subfigure{
\label{blend} 
\includegraphics[width=0.18\linewidth]{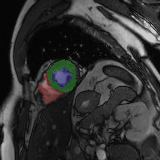}}
\subfigure{
\label{hsv} 
\includegraphics[width=0.18\linewidth]{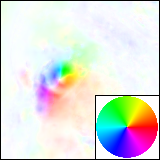}}
\caption{Examples of tracking results using the proposed method (MPN-C). From left column to right: ED frame, ES frame, warped frame from ED, overlay of ES frame and the warped mask, and estimated motion field using HSV color coding. The color coding wheel legend indicates the motion directions.}
\label{qualitative} 
\end{figure}

\begin{table}[bt]
\scriptsize
 \caption{Results of compared methods in terms of Dice coefficient on the ACDC and Kaggle dataset. Mean (standard deviation) is given.}
 \label{tab:acdc_dice_kaggle}
 \centering
 \begin{tabular}{ccccccc}
  \toprule
  \multirow{2}{*}{Method}   & \multicolumn{3}{c}{ACDC} & \multicolumn{3}{c}{Kaggle} \\
  \cmidrule(r){2-7} & LV &RV &MYO & LV&RV&MYO \\
  \midrule
 FFD & 0.893(0.077)  &0.850(0.124) & 0.793(0.080)  & 0.875(0.106)  &0.807(0.130) &0.793(0.113) \\
 Baseline & 0.885(0.111) & 0.872(0.113) & 0.830(0.057)& 0.867(0.104)  &0.816(0.128) & 0.800(0.108)\\
 MPN & 0.901(.097) & 0.880(0.108) & 0.847(0.048)& 0.883(0.095)  &0.820(0.127) & 0.820(0.102) \\
 MPN-S&   0.915(0.080) & 0.883(0.099) & {0.858(0.046)}& 0.893(0.080) & 0.819(0.127) & 0.833(0.084)\\
 MPN-C& \textbf{0.918(0.073)} & \textbf{0.885(0.096)} & \textbf{0.860(0.047))} & \textbf{0.896(0.075)} & \textbf{0.820(0.127)} & \textbf{0.837(0.077)} \\
 \hline
 \rule{0pt}{8pt} MPN-2& 0.913(0.084) & 0.887(0.102) & 0.861(0.044) & 0.898(0.080) & 0.822(0.128) & 0.842(0.083)\\
  \bottomrule
 \end{tabular}
\end{table}
\begin{table}[bt]
\scriptsize
 \caption{Results of compared methods in terms of Hausdorff distance on the ACDC and Kaggle dataset. Mean (standard deviation) millimeter is given.}
 \label{tab:hd_acdc_kaggle}
 \centering
 \begin{tabular}{ccccccc}
  \toprule
  \multirow{2}{*}{Method}   & \multicolumn{3}{c}{ACDC} & \multicolumn{3}{c}{Kaggle} \\
  \cmidrule(r){2-7} & LV &RV &MYO & LV&RV&MYO \\
  \midrule
  FFD & 5.99(2.53)  & 8.19(5.06) & 7.09(2.26)   &6.49(4.27) & 12.22(6.46) & 8.35(7.00) \\
  Baseline & 5.90(2.67) & 7.51(4.89) & 5.96(2.23)& 7.27(4.34) & 12.13(6.41) & 8.12(7.07)\\
 MPN & 5.51(2.49) & 7.26(4.91) & 5.59(2.09)& 6.60(4.19) & 12.04(6.47) & 7.72(7.03) \\
 MPN-S&  4.93(1.88) &7.22(4.77) & 5.38(1.90)& \textbf{6.28(3.95)} & \textbf{12.02(6.50)} & \textbf{7.61(6.86)}\\
 MPN-C& \textbf{4.82(1.78)} & \textbf{7.13(4.62)} & \textbf{5.41(1.72)} & 6.32(3.96) & 12.20(6.46) & 7.68(6.78) \\
 \hline
\rule{0pt}{8pt}  MPN-2& 4.95(2.03) & 7.15(4.85) & 5.36(1.85) & 6.19(3.95) & 11.99(6.52) & 7.46(6.91)\\
  \bottomrule
 \end{tabular}
\end{table}

\begin{table}[bt]
\scriptsize
 \caption{Results of compared methods in terms of average symmetrical surface distance on the ACDC and Kaggle dataset. Mean (standard deviation) millimeter is given.}
 \label{tab:assd_acdc_kaggle}
 \centering
 \begin{tabular}{ccccccc}
  \toprule
  \multirow{2}{*}{Method}   & \multicolumn{3}{c}{ACDC} & \multicolumn{3}{c}{Kaggle} \\
  \cmidrule(r){2-7} & LV &RV &MYO & LV&RV&MYO \\
  \midrule
  FFD & 1.96(1.05)  & 2.16(1.12)& 1.81(0.50)   & 2.24(2.00) &  2.72(1.48) & 1.95(1.12) \\
  Baseline & 1.98(1.20) & 1.85(1.01) & 1.61(0.57) & 2.44(1.84) &  2.60(1.45) & 1.89(1.07) \\
 MPN & 1.70(0.97) & 1.71(0.93) & 1.49(0.47) & 2.09(1.67) & 2.52(1.44) &1.76(1.04) \\
 MPN-S&  1.45(0.70) &1.65(0.84) & 1.40(0.42)& 1.86(1.34) & 2.48(1.41) &1.70(0.98)\\
 MPN-C& \textbf{1.38(0.59)} & \textbf{1.60(0.81)} & \textbf{1.38(0.40)} & \textbf{1.78(1.20)} & \textbf{2.47(1.40)} & \textbf{1.70(0.97)} \\
 \hline
 \rule{0pt}{8pt} MPN-2& 1.45(0.71) & 1.57(0.82) & 1.39(0.42) & 1.75(1.33) & 2.43(1.41) &1.62(0.98)\\
  \bottomrule
 \end{tabular}
\end{table}

\begin{table}[bt]
\scriptsize
 \caption{Endpoint error (EPE), its decomposition ($\vec{\epsilon}_{rr}$, $\vec{\epsilon}_{cc}$) in ACDC and Kaggle dataset. The keypoint tracking error is evaluated with ACDC only. Mean (standard deviation) millimeter is given.}
 \label{tab:epe_err}
 \centering
 \begin{tabular}{cccc p{1.6cm} ccc}
  \toprule
  \multirow{2}{*}{Method}   & \multicolumn{4}{c}{ACDC} & \multicolumn{3}{c}{Kaggle} \\
  \cmidrule(r){2-8} & EPE & $\vec{\epsilon}_{rr}$ & $\vec{\epsilon}_{cc}$& KPTE  & EPE & $\vec{\epsilon}_{rr}$ & $\vec{\epsilon}_{cc}$\\
  \midrule
  FFD & 2.70(1.51) & 1.62(1.01) & 1.44(1.14)  &  2.48(1.06) & 1.10(0.88) & 0.95(0.61)&2.17(1.70)\\
  Baseline & 1.82(1.48) & 1.08(0.77) & 1.17(1.15)  &  1.65(1.00) & 0.76(0.54) & 0.73(0.50) & 1.38(1.08)\\
 MPN & 1.79(1.45) & 0.95(0.70) & 1.17(1.16)  & 1.60(0.95) & 0.71(0.48) & 0.72(0.49) & 1.40(1.12)\\
 MPN-S&  1.73(1.41) &0.86(0.67) & \textbf{1.12(1.17)} & 1.56(0.89) &0.65(0.44) & 0.70(0.48)&1.29(1.05)\\
 MPN-C& \textbf{1.69(1.42)} & \textbf{0.84(0.65)} & 1.13(1.19) & \textbf{1.53(0.88)} & \textbf{0.64(0.42)} & \textbf{0.70(0.47)}& \textbf{1.27(1.02)}\\
 \hline
\rule{0pt}{8pt}  MPN-2& 1.71(1.49) & 0.92(0.81) & 1.11(1.25)  & 1.50(0.92) & 0.65(0.47) & 0.68(0.47)&1.33(1.16)\\
  \bottomrule
 \end{tabular}
\end{table}

We compared proposed methods: motion pyramid network (MPN), progressive motion compensation (MPN-2), teacher-student training (MPN-S), and cyclic knowledge distillation (MPN-C). Besides, we implemented a single-scale motion estimation network using the same network structure as MPN without the pyramids (baseline). Furthermore, we compared a conventional registration-based method: free form deformation (FFD) \cite{rueckert1999nonrigid}. All models are trained and tested on a Tesla V100 workstation. We evaluate all methods in 2 public datasets with multiple metrics. \textbf{ACDC \cite{bernard2018deep}} dataset is a short-axis CMR dataset collected from real clinical exams at a hospital in France. The total number of studies is 150, where 100 studies are the training set with expert manual segmentation annotations in end-diastole (ED) and end-systole (ES) phases. \textbf{Kaggle} \cite{kaggle} is composed of 1140 subjects with short-axis, 4- and 2-chamber long-axis CMR scans without segmentation masks. 
 
For the ACDC dataset, we make use of the training set, where 80 cases are used to train our model and 20 cases for testing. We first train a teacher model described in Section \ref{backbone}. Then we perform the cyclic training described in Section \ref{cycle}. Empirically we set the number of cycles to 2 since we find cycles more than 2 contribute little to the performance. For all the models, we use the same hyper-parameters. We use the Kaggle dataset only for testing where we randomly pick 130 patients from short-axis studies and manually segment the myocardium on 3 image pairs at the middle slice. 

Since there is no ground truth cardiac motion field from real CMR cine, we evaluated methods by comparing the warped mask using the estimated motion field with the reference mask.  We use standard segmentation evaluation metrics: Dice coefficients, Hausdorff distance (HD) and average symmetrical surface distance (ASSD). Similar approach is used in  \cite{qin2018joint,zheng2019explainable,krebs2019learning,yu2020foal}. To generate more samples for evaluation, for the ACDC dataset, we manually labeled 2 extra frames: $\frac{ED+ES}{2}$ and $\frac{ED+ES}{2} + 1$, between the ED and the ES frame. For each test sample, we evaluate 3 pairs: $\{ED\rightarrow ES, ED\rightarrow \frac{ED+ES}{2}, ED\rightarrow \frac{ED+ES}{2} + 1\}$. We select the middle slice of each study to avert the impact of the out-of-plane motion. Table \ref{tab:acdc_dice_kaggle}, \ref{tab:hd_acdc_kaggle}, \ref{tab:assd_acdc_kaggle} show the compared results in terms of Dice, HD and ASSD in dataset ADDC and Kaggle. Results of MPN-2 are separated since it uses multi-step compensation while others are single-step. Two example results using MPN-C are depicted in Fig.~\ref{qualitative}.

\subsection{Endpoint error and its decomposition}
\label{sec:decomp}
In the field of motion estimation, the most commonly used evaluation metric is endpoint error (EPE) \cite{meister2018unflow,sun2018pwc}. For 2 frames $I_A$ and $I_B$, suppose the estimated flow is $\hat f_{AB}$ and the ground truth flow is $f_{AB}$, $EPE = ||\hat f_{AB} - f_{AB}||_2$.

However, the computation of EPE requires the known ground truth motion field, which is difficult to acquire for the cardiac motion estimation task. We use a different model trained with the same dataset to estimate the motion field and treat it as ground truth. Specifically, we use a single scale model similar to the motion estimation branch described in \cite{qin2018joint}. Using the generated motion field to warp the image, we can get a synthesized image pair to evaluate the EPE. 

EPE measures the magnitude of the error vectors, which is sufficient for the general motion estimation tasks. The goal of cardiac motion estimation is different: motion is used to calculate clinical indices like strain along radial and circumferential directions of the myocardium, which makes it necessary to decompose the error in those directions. We propose a method to decompose the endpoint error in a clinically meaningful way. First, compute the center of the myocardium region $ x_c = \frac{1}{N}\sum_{ x_i \in {myo}}^{N} x_i$. The radial direction of every point within myocardium can be computed as $\vec{d}(x_i) = x_i-x_c$ and then normalized to unit vector. 
Endpoint error vector $\vec{e}_i$ at $x_i$ is $\vec{e}_i = f(x_i) - \hat{f}(x_i)$ which is decomposed along radial ($\vec{\epsilon}_{rr}$) and circumferential ($\vec{\epsilon}_{cc}$) directions as: $\vec{\epsilon}_{rr}^{(i)} = \vec{e_i} \cdot \vec{d}(x_i)$, $\vec{\epsilon}_{cc}^{(i)} = \vec{e}_i - \vec{\epsilon}_{rr}^{(i)}$. Experiment results are  presented in Table \ref{tab:epe_err}.

\subsection{Key point tracking error}
Quantitative plots like the bullseye plot \cite{young2009computational} are widely used in clinical applications, which requires an accurate partition of the myocardium. The insertion points of the left and the right ventricles are key points for the partition \cite{berman2004prognostic}. We propose Key Point Tracking Error (KPTE) to measure the error of these landmarks. For $N$ predicted key points $\{\hat x_i\}$ and the ground truth locations $\{x_i\}$, $KPTE = \frac{1}{N}\sum^N_{i} ||x_i - \hat x_i||$.
We first manually label the key points on the ED frame and these points are warped with synthesized motion as the ground truth locations on the following frames, using the same method in Section \ref{sec:decomp}. Then we use the estimated motion to predict the locations of the key points and compute KPTE. The evaluation results are shown in Table \ref{tab:epe_err}.

\subsection{Generalization study}
The proposed cyclic knowledge distillation training method with progressive motion compensation is not limited to certain neural network structure. We applied this to the recently proposed motion estimation model Apparentflow \cite{zheng2019explainable}. In contrast to \cite{zheng2019explainable}, we only train their model with the self-supervision loss regardless of segmentation mask supervision. Also, we extended this work using our methods in Section \ref{cycle} and \ref{compensation}. We coined them as Apparentflow-C and Apparentflow-2 respectively. The compared results are presented in Table \ref{tab:generalize}.

\begin{table}[bt]
\scriptsize
 \caption{Performance measures for the Apparentflow and applying the proposed cyclic knowledge distillation and progressive motion compensation to it on the ACDC dataset. Mean (standard deviation) is given for the Dice coefficient.}
 \label{tab:generalize}
 \centering
 \begin{tabular}{cccc}
  \toprule
  Method & LV& RV & MYO \\
  \midrule
 Apparentflow & 0.880(0.111) & 0.870(0.110) & 0.811(0.070) \\
Apparentflow-C & \textbf{0.902(0.098)} & \textbf{0.883(0.095)} & \textbf{0.837(0.051)}  \\
\hline
Apparentflow-2 & 0.901(0.105)&0.885(0.094)&0.841(0.049) \\ 
  \bottomrule
 \end{tabular}
\end{table}

\subsection{Results discussion}
 From Table \ref{tab:acdc_dice_kaggle}, \ref{tab:hd_acdc_kaggle} and \ref{tab:assd_acdc_kaggle}, we observe that MPN outperforms deformable registration and baseline model in terms of Dice, HD and ASSD. Moreover, the utilization of motion compensation (MPN-2) improves the MPN's performance. The teacher-student model (MPN-S) has comparable accuracy as MPN-2, which demonstrates knowledge distillation helps parameter searching during training. The cyclic training strategy (MPN-C) further pushes the performance. From Table \ref{tab:epe_err}, we have the same conclusion in the synthetic dataset in terms of endpoint error. Radial tracking is more accurate than circumferential tracking in light of $\vec{\epsilon}_{rr}$ and $\vec{\epsilon}_{cc}$ errors. This complies with our knowledge that circumferential tracking of circular shape object from 2D in-plane images is a challenging task. Results in Table \ref{tab:generalize} demonstrates our methods are not limited to our own network structure but also applicable to other motion models (i.e. Apparentflow \cite{zheng2019explainable}). On average, FFD takes 748 milliseconds (ms) to register one frame while our proposed model (MPN-C) needs 26 ms for tracking one frame. 
\section{Conclusion}

In summary, we proposed a novel deep neural network model that exploits multi-scale supervision. Specifically, we utilize a progressive motion compensation strategy to overcome the limitation of motion smoothness constraints. We developed a new training strategy, the proposed cyclic knowledge distillation, which helps the learner gain inference capability of several progressive motion compensation steps. Also, we proposed and evaluated two novel evaluation metrics for CMR-FT task: error decomposition and key point tracking error in addition to the Dice coefficient and boundary distance error. In the future, we plan to apply our method to a clinical study for the short-axis CMR-FT task.

%
%
%
\bibliographystyle{splncs04}
\bibliography{ref}

%
\end{document}